\documentclass[10pt,conference]{IEEEtran}                                                          

\IEEEoverridecommandlockouts                              

\usepackage{booktabs} 
\usepackage{tabularx}
\usepackage{multirow}
\usepackage{float}
\usepackage{listings}
\usepackage{color}
\usepackage{microtype}
\usepackage{url}

\usepackage{xspace}
\usepackage{tikz}
\usepackage{pgfplots}
\usepackage{adjustbox}
\usepackage{array}

\makeatletter
\newcommand{\thickhline}{%
    \noalign {\ifnum 0=`}\fi \hrule height 0.9pt
    \futurelet \reserved@a \@xhline
}
\makeatother

\begin{document}
\title{Evaluating Manual Intervention to Address the Challenges of Bug Finding with KLEE}

\newcommand{\KLEE}{\texttt{KLEE}\xspace}
\newcommand{\KLEES}{\texttt{KLEE}'s\xspace}
\newcommand{\KLEEs}{\KLEES}
\newcommand{\ic}[1] {\texttt{#1}}

\author{\IEEEauthorblockN{John Galea,
Sean Heelan, Daniel Neville and
Daniel Kroening}
\IEEEauthorblockA{University of Oxford\\
\{john.galea, sean.heelan, daniel.neville, kroening@cs.ox.ac.uk\}}}
\maketitle

\maketitle

\begin{abstract}
Symbolic execution has shown its ability to find security-relevant flaws in software, but faces significant scalability challenges. There is a commonly held belief that manual intervention by an expert can help alleviate these limiting factors. However, there has been little formal investigation of this idea. In this paper, we present our experiences applying the KLEE symbolic execution engine to a new bug corpus, and of using manual intervention to alleviate the issues encountered. Our contributions are (1)~\ic{Hemiptera}, a novel corpus of over 130 bugs in real world software, (2)~a comprehensive evaluation of the \KLEE symbolic execution engine on \ic{Hemiptera} with a categorisation of frequently occurring software patterns that are problematic for symbolic execution, and (3)~an evaluation of manual mitigations aimed at addressing the underlying issues of symbolic execution. Our experience shows that manual intervention can increase both code coverage and bug detection in many situations. It is not a silver bullet however, and we discuss its limitations and the challenges encountered. 
\end{abstract}

\section{Introduction}
Symbolic execution is a flexible approach to program analysis which can be embedded in various approaches to program analysis. Applications include test case generation \cite{godefroid2012sage}, static bug finding \cite{cadar2008klee}, patch validation \cite{ramos2015under}, deobfuscation \cite{salwan2017deobfuscation} and exploit generation \cite{cha2012unleashing}. However, when applying symbolic execution on a new application it is common to encounter significant problems. In this paper we document our experiences using KLEE to perform bug detection across a diverse set of programs, and of using manual intervention to alleviate the issues encountered.

The term \emph{symbolic execution} refers to analysis systems with a broad range of capabilities and approaches. In this paper, we focus on a path-wise forwards analysis that uses symbolic variables for the inputs from the environment. We exclude from consideration systems that utilise concrete inputs in order to produce a trace that is then executed symbolically (e.g., concolic execution~\cite{godefroid2012sage}). We also exclude systems that utilise state merging in order to produce states that represent multiple paths~ \cite{avgerinos2014enhancing, kuznetsov2012efficient}. The rationale for this focus is that the majority of maintained open source implementations\footnote{\url{https://klee.github.io}}\footnote{\url{https://angr.io}}\footnote{\url{https://github.com/trailofbits/manticore}} do not implement either of these features. While state merging and concolic execution \emph{may} resolve some issues we discuss, at the present time, these features are unavailable to a user who wants to utilise symbolic execution via the existing tools.

We furthermore focus on the \KLEE symbolic execution engine~\cite{cadar2008klee} in our experimental work. \KLEE is a mature, maintained symbolic execution engine and has served as the basis for a large number of academic publications\footnote{\url{https://klee.github.io/publications/}} from a diverse set of research groups.

Generally, existing approaches that attempt to address the scalability issues faced by symbolic execution are aligned in two complementary directions.  Firstly, there are automated solutions, for example state merging~\cite{kuznetsov2012efficient,avgerinos2014enhancing} and compositional testing~\cite{anipaper}.  Secondly, there are manual solutions, such as the introduction of logical assumptions~\cite{kleeassume2017team,vanegue2013staticcheckers}, where a domain expert assists symbolic execution in order to improve effectiveness.



While path explosion can be demonstrated with a very simple program, namely, a loop that contains a branch, detailed case-studies of problematic patterns that occur in real-world software are lacking in the literature. Similarly, while many members of the research community expect that manual intervention can deal with problematic patterns in the software under test, there are few experience reports on how feasible this approach is across diverse applications and how successful, or not, it can be. We aim to address both of these issues.

To generate a baseline dataset we use \KLEE~\cite{cadar2008klee} to analyse a set of real-world applications with the goal of discovering security-relevant bugs. We~document the limiting factors we observe and categorise the root causes of these issues. To mitigate the challenges encountered, we employ existing techniques that require manual intervention, which often involve trading soundness or completeness for scalability. Exemplars are the manual insertion of logical constraints, early state termination and the use of driver programs. Finally, we evaluate the changes in code coverage and bug detection that result from these interventions and discuss their successes and limitations. 


We view our research as orthogonal to work on automated solutions to the categories of problems that we describe. An automated solution that effectively manages one limitation still often leaves the symbolic execution engine struggling due to other issues left unaddressed. This hampers adoption of symbolic methods. Thus, it is necessary to also understand the outstanding issues that are significant when analysing real programs, and how these issues can be alleviated via manual intervention. 

To support our research, we have compiled \ic{Hemiptera}, a substantial bug suite containing 133 genuine bugs found in open-source applications. The bugs are individually labelled with commit IDs identifying a software version they are present in, and test cases that trigger the bugs are provided. A substantial amount of previous work uses synthetic datasets~\cite{stephens2016driller}, a very limited number of real-world programs~\cite{kuznetsov2010ddt, wang2009intscope} or real-world programs where a lower bound for the number of bugs that \emph{should} be detectable is not known~\cite{avgerinos2014enhancing, cadar2008klee}. \ic{Hemiptera} is shared openly with the community to support further research.

\subsection{Contributions}

This paper provides three main contributions.
\begin{enumerate}
    \item  \ic{Hemiptera}, a bug suite containing over 130 genuine bugs in open-source programs, with test cases, and labelled with their associated commit IDs.
    \item A categorisation of code constructs that pose severe challenges to symbolic execution via a large-scale investigation of more than 24 program versions, across eight programs, using \KLEE. 
    \item An evaluation of the feasibility and effectiveness of manual intervention in addressing state space explosion within symbolic execution.
\end{enumerate}

We provide all raw data, scripts and source within the artefacts\footnote{Via GitHub, this link has been removed to anonymise the paper for review}. 

\subsection{Organisation of Paper}

The rest of the paper is organised as follows: Section~2 delves into \ic{Hemiptera},  Section~3 presents an evaluation of \KLEE on \ic{Hemiptera},  Section~4 presents an in-depth exploration of the challenges encountered by symbolic execution in the evaluation from Section~3, and the mitigations we employ to overcome these challenges,  Section~5 presents an evaluation of the success of the mitigations described in Section~4, Section~6 lists potential threats to validity, Section 7 discusses related work, and Section~8 concludes the paper.

\section{The Hemiptera Bug Suite}
\texttt{Hemiptera}\footnote{Our bug suite is named after the order of bugs \ic{Hemiptera}, meaning ``True Bugs''.} is  a large corpus of patched bugs that have been present in eight major real-world applications. In total, the corpus contains 133 bugs, many of which are security relevant and could be leveraged to gain malicious code execution. The types of bugs present in \ic{Hemiptera} include invalid memory accesses, failed assertions, and division by zero exceptions. The applications considered are all open-source and written in~C. 

While some existing bug suites~\cite{lu2005bugbench} do consider real-world applications, they are often limited due to small bug counts.  Meanwhile, others~\cite{boland2012juliet, ku2007buffer, kratkiewicz2005using} provide a large number of bugs but are synthetic. For instance, Dolan-Gavitt et al.~\cite{Dolan-Gavitt2016a} have proposed LAVA, a tool capable of automatically injecting artificial bugs into software. Although such an approach facilitates the creation of a large bug corpus, it still threatens the validity of the claims that might be made about real-world code when used as the sole evaluation corpus for an analysis. The DARPA CGC corpus also faces similar issues\footnote{\url{http://archive.darpa.mil/cybergrandchallenge/tech.html}}, since the programs are synthetic. 

\begin{table*}[t]
\centering
\caption{Summary of \texttt{Hemiptera}'s composition.}
\label{table-bug}
\begin{tabular}{|c|c|c|c|c|c|c|}
\hline
\textbf{Project} & \textbf{\begin{tabular}[c]{@{}c@{}}\# Invalid \\ Accesses\end{tabular}} & \textbf{\begin{tabular}[c]{@{}c@{}}\# Failed \\ Asserts\end{tabular}} & \textbf{\begin{tabular}[c]{@{}c@{}}\# Div \\ by Zeros\end{tabular}} & \textbf{\begin{tabular}[c]{@{}c@{}}\# Dang\\ Ptr. Use\end{tabular}} & \textbf{\begin{tabular}[c]{@{}c@{}}Size of \\ Min-Sets\end{tabular}} & \textbf{\begin{tabular}[c]{@{}c@{}}Avg. lines \\ of C Code\end{tabular}} \\ \hline
\ic{JasPer}           & 11                                                                         & 6                                                                        & 2                                                                           & 0                                                                          & 2                                                                    &    25,891                                                                    \\ \hline
\ic{LibTIFF}          & 36                                                                         & 1                                                                        & 4                                                                           & 0                                                                          & 5                                                                    & 56,573                                                                        \\ \hline
\ic{tcpdump}          & 10                                                                         & 0                                                                        & 0                                                                           & 0                                                                          & 2                                                                    & 47,782                                                                       \\ \hline
\ic{libjpeg-turbo}    & 1                                                                          & 0                                                                        & 1                                                                           & 0                                                                          & 1                                                                    &    25,920                                                                    \\ \hline
\ic{zlib}             & 2                                                                          & 0                                                                        & 0                                                                           & 1                                                                          & 2                                                                    &   7,835                                                                   \\ \hline
\ic{file}             & 5                                                                          & 1                                                                        & 1                                                                           & 0                                                                          & 3                                                                    &   12,597                                                                     \\ \hline
\ic{w3m}           & 44                                                                          & 0                                                                        & 0                                                                           & 2                                                                          & 3                                                                    &   53,338                                                                   \\ \hline

\ic{FLAC}             & 5                                                                          & 0                                                                        & 0                                                                           & 0                                                                          & 1                                                                    &   46,114                                                                   \\ \hline \hline

Total             & 114                                                                          & 8                                                                        & 8                                                                           & 3                                                                          & 19                                                                    &   34,506                                                                   \\ \hline

\end{tabular}
\end{table*}


\paragraph{Composition.}
Table \ref{table-bug} summarises the composition of \texttt{Hemiptera}. Several of the bugs contained in \ic{Hemiptera} are security relevant, and could have been leveraged to gain malicious code execution. Where applicable, \ic{Hemiptera} provides the CVE ID that is associated with a given bug. 

For each bug in the corpus, we recorded the commit ID of the patch as well as the commit ID of the immediate unfixed predecessor (the parent ID).  The inclusion of every unfixed version, for each bug, would result in an impractically high number of application versions to analyse.  Consequently, for each application, \texttt{Hemiptera} provides the minimal set of commit IDs (referred to as a \textit{Min-Set}) such that all of the bugs in the corpus that pertain to an application are present. We calculated the average line counts of the versions in each \textit{Min-Set}. Overall, the corpus considers reasonably large applications, ranging from 7,835 (\ic{zlib}) to 56,573 (\ic{LibTIFF}) lines of C code. 

\paragraph{Selection Criteria.}


We selected applications for inclusion in \texttt{Hemiptera} based on four major criteria:
%
\begin{enumerate}
 \item The applications are similar to those considered suitable for analysis by state of the art symbolic execution tools. Examples are file parsers, utility libraries and system tools.
 \item If an application contained bugs, such bugs would be likely to have security implications.
 \item The application's project must be open-source.
 \item The project must have a publicly accessible version control system (e.g.~Git), such that changes to files are tracked, and we can therefore identify the buggy and patched versions for each bug.
\end{enumerate}


 
\paragraph{Construction Methodology.}
 
\begin{figure}[t!]
  \centering
  \includegraphics[width=0.45\textwidth]{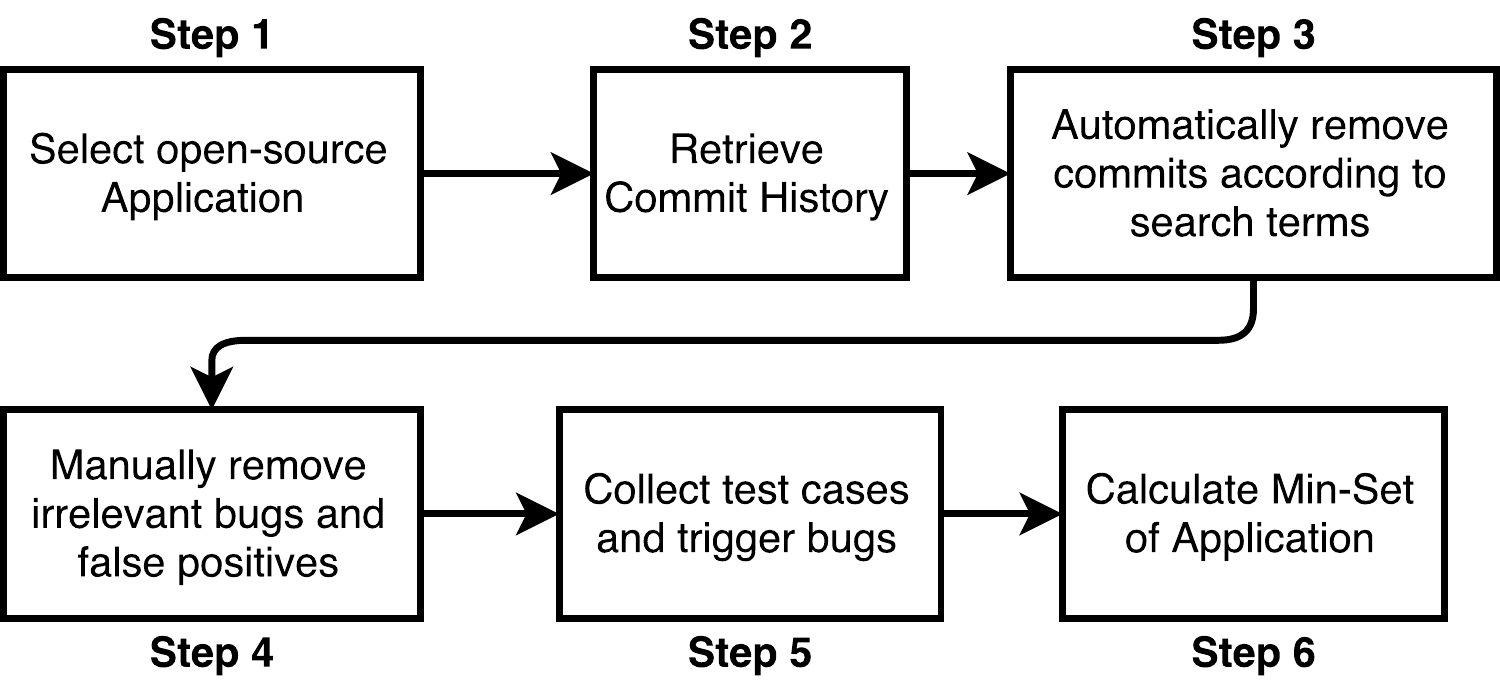}
  \caption{Methodology of building \texttt{Hemiptera}.}
  \label{fig:hemiptera_fc}
\end{figure}
 
Figure \ref{fig:hemiptera_fc} illustrates the methodology of \texttt{Hemiptera}'s construction. Once an application is selected, we run a set of scripts to facilitate the inspection of change logs. The scripts automatically retrieve the commit history of the master branch of the application, and select commits that have messages that contain relevant terms. These terms include ``bug", ``cve", ``zero", ``overflow" and ``crash". For all selected commits, the scripts extract their date, hash ID and parent ID. The process continues with manual analysis, whereby the resulting commits are inspected to check whether they are likely to be patches for security relevant bugs. Bugs that seem to be security irrelevant, such as memory leaks and syntax errors, or are very platform dependent, are discarded.

The presence of these bugs is then confirmed by searching for an input which triggers it. Test cases that successfully trigger the bugs are retrieved by inspecting bug trackers, test folders, and online repositories, such as Exploit-DB\footnote{\url{https://www.exploit-db.com/}}. Failure to find a test case that successfully triggers a bug results in the bug being omitted from the corpus. 

The process then takes an incremental approach in order to derive the \textit{Min-Set} of the application. It begins by first including the commit ID of earliest buggy version to the set. We then run the version against the collected test cases iteratively, in an order that is sorted by the date of fix of the bugs. In the event that a test case fails to reproduce its bug, the commit ID of the version that successfully triggers the bug is included in the \textit{Min-Set}. In turn, the version associated with the newly inserted commit ID is considered for checking the remaining test cases. This process continues until all bugs are triggered successfully. 

Although the process of building \texttt{Hemiptera} has been partially automated through scripting, it still incurred a significant amount of manual effort. We believe that dealing with this challenge is worthwhile to reduce threats to the validity of our results, as well as contribute a useful bug corpus to the community.

\section{Evaluation of the State of the Art}
\label{sec:initial_results}
To generate a dataset to categorise challenges in real-world symbolic execution, we ran \KLEE on 24 different program versions, across eight programs, as provided by \ic{Hemiptera}. The configuration for \KLEE as recommended by its authors\footnote{\url{https://klee.github.io/docs/coreutils-experiments/}} was used, combined with our own \KLEE modifications as detailed in the next section.

\subsection{Experiment}

We used \KLEE version 1.3.0 with two search strategies, \emph{covnew} and \emph{random-path}\footnote{\url{https://klee.github.io/docs/options/}}. The \emph{covnew} state selection algorithm selects the state it believes to be closest to uncovered code, while the \emph{random-path} algorithm selects a state by randomly traversing the tree of all paths taken by all live states until a state representing the head of a path is reached. 

We also tested a modification for \KLEE, which was introduced to resolve an issue we encountered with state selection: When using the \emph{batching searcher}, if a conditional branch based on symbolic data is encountered, where both the taken and fall-through cases are feasible, \KLEE will always schedule the state corresponding to the branch taken case. On several targets, this often results in failure to analyse the remaining code in functions that have a branch towards a \ic{return} early in their body. Consequently, we implemented a mechanism that randomly selects between the two states of a branch, post forking. We refer to this feature as \emph{rspf} (random state, post-fork) in the remainder of this section.

As well as using \emph{covnew} and \emph{random-path} together, we also ran \KLEE against each program with only \emph{random-path}, and with a combination of \emph{covnew}, \emph{random-path}, and \emph{rspf}. All target applications were compiled to LLVM bitcode using \ic{wllvm}\footnote{\url{https://github.com/travitch/whole-program-llvm}}. Optional third-party libraries were not linked, unless they were required for a bug found in \ic{Hemiptera}. As \ic{libtiff} comes with a significant number of driver applications, we focused specifically on \ic{tiff2pdf}, which triggers the highest proportion of its bugs. Each run was one hour long, and each experiment was executed three times to increase reliability.  We ran our experiments on an Intel Xeon X5667 at 3.07\,GHz with 24\,GB of RAM available to each process.

\subsection{Results}

Table~\ref{tab:default} quantifies the effectiveness of \KLEE.  We present averages over all runs within the table. The performance appears to be generally poor, finding very few of the known bugs in \texttt{Hemiptera}. \KLEE failed to find bugs in most program versions.   In general, code coverage appears to be low in absolute terms, but without any external baseline to compare against, it is difficult to determine what proportion of ``interesting'' code this corresponds to. However, considering most bugs are not found, it is clear that a substantial amount of ``interesting'' code is not reached.

When using \emph{covnew} combined with \emph{random-path}, \KLEE found bugs in six different program versions, while \emph{rspf} improved bug detection for one program version. By contrast, \emph{random-path} finds significantly fewer bugs (a total of six bugs compared to a total of 109 bugs for \emph{covnew} + \emph{random-path}). However, on some programs (e.g.\ \ic{imginfo}, \ic{zlib} and \ic{file}), \emph{random-path} achieves \emph{higher} coverage, despite the fact that \emph{covnew} has been designed specifically to maximise coverage. Furthermore, if we exclude \ic{tcpdump}, then \emph{random-path} has nearly equivalent code coverage and bug finding performance to the coverage-driven runs. 

\begin{table*}[ht]
\centering
\caption{Bug count and coverage results attained by \KLEE on real-world programs. \textit{H. Bugs} represents the detected bugs which are also presented in \ic{Hemiptera}.}
\label{tab:default}
\begin{tabular}{|l|l|r|r|r|r|r|r|r|r|r|r|}
\hline
\multicolumn{1}{|c|}
{\multirow{3}{*}{\textbf{Program}}} & \multicolumn{1}{c|}{\multirow{3}{*}{\textbf{Commit}\footnotemark}} & \multicolumn{3}{c|}{\textbf{\begin{tabular}[c]{@{}c@{}}\emph{covnew} \& \\ \emph{random-path}\end{tabular}}} & \multicolumn{3}{c|}{\textbf{\emph{random-path}}} & \multicolumn{3}{c|}{\textbf{\begin{tabular}[c]{@{}c@{}}\emph{covnew}, \\ \emph{random-path} \& \\ \emph{rspf}\end{tabular}}} & \multicolumn{1}{c|}{\multirow{3}{*}{\textbf{\begin{tabular}[c]{@{}c@{}}Known\\Bugs in\\Hemiptera\footnotemark \end{tabular}}}} \\ \cline{3-11}
\multicolumn{1}{|c|}{}                                 & \multicolumn{1}{c|}{}                                 & \textbf{Bugs}    & \textbf{H. Bugs}       & \textbf{ICov\footnotemark}         & \textbf{Bugs}  & \textbf{H. Bugs} & \textbf{ICov}      & \textbf{Bugs}   &  \textbf{H. Bugs}   & \textbf{ICov}      & \multicolumn{1}{c|}{}                                                                                   \\ \hline
tcpdump                                                & 17a3c288*                                          & 32 & -                            & 4.94\%                & 0& -                         & 3.14\%             & 28 & -                       & 4.23\%             & 0                                                                                                       \\ \hline
tcpdump                                                & a9e4211                                               & 40  & 1                          & 4.91\%                & 0 & 0                         & 3.39\%             & 28 & 2                       & 4.20\%             & 9                                                                                                       \\ \hline
tcpdump                                                & 8322d3a                                               & 31 & 0                            & 4.61\%                & 0 & 0                        & 3.14\%             & 23 & 0                        & 4.40\%             & 1                                                                                                       \\ \thickhline
w3m                                                    & 1ac245b*                                              & 0 & -                            & 9.01\%                & 0 & -                        & 8.77\%             & 0 & -                        & 9.15\%             & 0                                                                                                       \\ \hline
w3m                                                    & 02ba3d6                                               & 4 & 0                             & 8.96\%                & 3 & 0                        & 8.75\%             & 4 & 1                         & 9.07\%             & 9                                                                                                       \\ \hline
w3m                                                    & 06caca1                                               & 0 & 0                            & 8.90\%                & 0 & 0                         & 8.74\%             & 0 & 0                         & 9.01\%             & 15                                                                                                      \\ \hline
w3m                                                    & a56a8ef                                               & 0  & 0                           & 8.96\%                & 0 & 0                         & 8.75\%             & 0 & 0                         & 9.06\%             & 11                                                                                                      \\ \thickhline
libjpeg-turbo                                                & a0b7de9*                                              & 0  & -                            & 23.10\%               & 0 & -                         & 22.72\%            & 0 & -                          & 20.61\%            & 0                                                                                                       \\ \hline
libjpeg-turbo                                                & 3091354                                               & 1 & 0                             & 15.01\%               & 1 & 0                         & 14.98\%            & 1 & 0                         & 15.38\%            & 2                                                                                                       \\ \thickhline
imginfo                                                & 4212e7e*                                              & 0 & -                             & 10.36\%               & 0 & -                         & 14.68\%            & 0 & -                         & 9.93\%             & 0                                                                                                       \\ \hline
imginfo                                                & b702259                                               & 1 & 1                             & 10.39\%               & 2 & 1                         & 12.41\%            & 1  & 1                        & 11.08\%            & 18                                                                                                      \\ \thickhline
jasper                                                 & 4212e7e*                                              & 0 & -                              & 12.96\%               & 0 & -                          & 13.26\%            & 0 & -                          & 13.93\%            & 0                                                                                                       \\ \hline

jasper                                                 & ed355a6                                               & 0 & 0                             & 13.47\%               & 0 & 0                       & 13.75\%            & 0 & 0                          & 12.04\%            & 2                                                                                                       \\ \thickhline

file                                                   & df74b09*                                              & 0 & -                            & 12.61\%               & 0 & -                        & 12.75\%            & 0 & -                         & 8.82\%             & 0                                                                                                      \\ \hline
file                                                   & b6e8437                                               & 0 & 0                            & 13.83\%               & 0 & 0                        & 13.90\%            & 0 & 0                        & 13.94\%            & 6                                                                                                       \\ \hline
file                                                   & 4a51454                                               & 0 & 0                             & 17.11\%               & 0 & 0                        & 17.73\%            & 0 & 0                        & 17.70\%            & 2                                                                                                       \\ \hline
file                                                   & 7445748                                               & 0 & 0                             & 15.63\%               & 0 & 0                        & 17.80\%            & 0 & 0                        & 17.04\%            & 1                                                                                                       \\ \thickhline
zlib                                                   & cacf7f1d*                                             & 0 & -                            & 26.70\%               & 0 & -                        & 27.38\%            & 0 & -                        & 25.08\%            & 0
 \\ \hline

zlib                                                   & 7c2a874                                               & 0 & 0                            & 26.56\%               & 0 & 0                         & 29.08\%            & 0 & 0                         & 24.45\%            & 2                                                                                                       \\ \hline

zlib                                                   & 14763ac                                               & 0 & 0                            & 31.44\%               & 0  & 0                       & 31.65\%            & 0 & 0                        & 32.12\%            & 1                                                                                                       \\ \thickhline
tiff2pdf                                               & d57ccfc9*                                             & 0 & -                            & 8.48\%                & 0 & -                         & 8.37\%             & 0 & -                         & 8.27\%             & 0                                                                                                       \\ \hline

tiff2pdf                                               & f64949a                                               & 0 & 0                            & 8.42\%                & 0 & 0                         & 8.47\%             & 0 & 0                         & 8.47\%             & 3                                                                                                       \\ \thickhline
flac                                                   & d2cb0d1*                                              & 0 & -                            & 12.48\%               & 0 & -                        & 12.16\%            & 0 & -                        & 13.77\%            & 0                                                                                                       \\ \hline
flac                                                   & d8d1717                                               & 0  & 0                           & 12.29\%               & 0 & 0                        & 12.11\%            & 1 & 1                         & 14.23\%            & 5                                                                                                       \\ \hline

\end{tabular}
\end{table*}

\addtocounter{footnote}{-3}

\stepcounter{footnote}\footnotetext{An asterisk on the commit number represents that the head version of a project was considered.}
\stepcounter{footnote}\footnotetext{It is unknown how many bugs exist in the latest version of a program.}
\stepcounter{footnote}\footnotetext{\ic{ICov} refers to the instruction coverage, which is calculated over all library code, even if it was unreachable.}

Several bugs found in tcpdump were unknown and present on the most recent version at the time of conducting our experiments. Therefore they were not included in Hemiptera, as the methodology of constructing our corpus requires that included bugs are patched.

One possible cause why \emph{covnew} performs worse than \emph{random-path}, in certain cases, is \emph{covnew}'s algorithm for calculating proximity to uncovered code. This algorithm is static, path-insensitive and over-approximates the points-to sets for function pointers. We discovered that \emph{covnew} frequently schedules states forked from the same location repeatedly, as it erroneously believes that a state at that location is close to uncovered code, and can reach it. \KLEE then runs the scheduled state but, importantly, never succeeds in actually reaching the code it thought it was close to. The algorithm does not react to this failure, e.g., by penalising the starting location, and therefore \emph{covnew} picks another state from \emph{exactly} the same initial location, only to fail again with high likelihood.

\pgfplotsset{compat=newest}

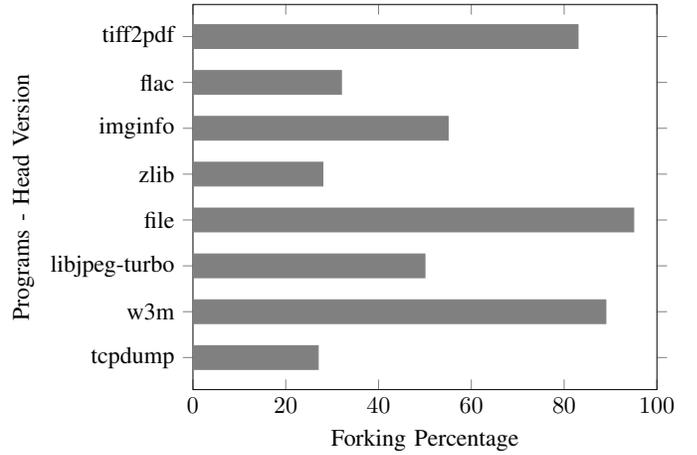
\begin{figure}[t!]
\centering

\begin{tikzpicture}[scale=0.9, transform shape]
\begin{axis}[
    ytick=data,
    xmin=0,
    xmax=100,
    xbar,
    ylabel={Programs - Head Version},
    xlabel={Forking Percentage},
    symbolic y coords={
    tcpdump,
    w3m, 
    libjpeg-turbo, 
    file, 
    zlib, 
    imginfo,
    flac,
    tiff2pdf},]
\addplot[gray,fill=gray] coordinates {
(27,tcpdump) 
(89,w3m) 
(50,libjpeg-turbo) 
(95,file) 
(28,zlib)
(55,imginfo) 
(32,flac) 
(83,tiff2pdf)
};
\end{axis}
\end{tikzpicture}

\caption{The percentage of state forks produced by the program location responsible for most state forks, per application. A single program location can often be responsible for the consumption of an inordinate amount of the analysis engine's resources.}

\label{fig:foking_percentage_chart}
\end{figure}






One notable observation is \KLEES vulnerability to code constructs that are similar to canonical examples of the path explosion problem. In Figure~\ref{fig:foking_percentage_chart} we can see that a single program location is often responsible for the majority of states created across the applications in \ic{Hemiptera}. Upon investigation, most of these locations are branches embedded in a straightforward loop over input.

It is clear that \emph{no} search strategy is particularly effective on all programs. On many applications in \ic{Hemiptera}, \KLEE struggles to find bugs and attain high code coverage. Moreover, whatever problems \KLEE is encountering are sufficiently severe to diminish the benefit of the coverage-driven search strategies in all but one application.   

\section{Categorisation of Challenges and Manual Interventions}
\label{sec:challenges}

In this section we present a categorisation of the main issues \KLEE encountered during the analysis of the applications in \ic{Hemiptera}, using the dataset generated via the experimental runs outlined in Section~\ref{sec:initial_results}. The categories cover the most significant issues encountered and we hypothesise that they are likely to generalise to targets not only in \ic{Hemiptera}, but within the realm of applications one might reasonably expect symbolic execution to be able to analyse.

We also offer mitigations for these issues, based on manual intervention, where applicable. Manual intervention is often proposed as a means by which one can alleviate the issues encountered by symbolic execution. However, there are few experience reports documenting its effectiveness on benchmarks like \ic{Hemiptera}. 

To investigate the challenges, we use the following information: (1)~runtime analysis data as generated by \KLEE, from which we can determine the program locations responsible for generating forks in the symbolic state and solver queries, (2)~data provided by \KLEEs existing analysis tools, such as \ic{klee-stats} and \ic{klee-replay}, (3)~logs of the scheduler's decision process\footnote{We extended \KLEEs logging in a number of different ways to allow us to understand what states were being scheduled and why, as well as what code they were covering.}, (4)~data on the final location for states still live at the end of the experiment\footnote{We also added logging of the states which were still alive at the end of each experiment, in order to determine what locations in the target states were blocked at.}, and (5)~warnings from \KLEE.

An analysis run is often hindered by a single dominant challenge.  Therefore, we have followed an iterative approach when introducing our modifications. We run each target for an hour, analyse the information mentioned above, introduce a small number of changes, and then repeat the process. We~stopped iteration when there was no longer a clear significant challenge that we can address using manual intervention.

\subsection{State-Space Explosion of Semantically Similar States}
\label{sec:challenge_state_space_explosion}




Across all targets, we encountered constructs which resulted in state space explosion. Often, the path constraints of the resulting states did not differ in a manner which would have a significant impact on future branches. We deem such states as \emph{semantically similar}.

\subsubsection{State Forking in Avoidable Code}

One class of functions which fit this category, and commonly produce semantically similar states, are those related to the formatting and printing of data for output. For example, \ic{uclibc}'s \ic{printf}\footnote{\ic{klee-uclibc/libc/stdio/printf.c}}, and its called functions, contain branches which will produce new states if the arguments to \ic{printf} are symbolic. However, the resulting states are semantically similar as they only pertain to the printing of data. When analysing \ic{tcpdump}, the functionality related to the formatting and printing of data is responsible for 61\% of the states forked, while on \ic{libtiff} similar functionality was responsible for 10\% of solver time. 

Such state forking can be resolved by introducing a \emph{low fidelity model} of the functions in question. A low fidelity model is a significantly less complex replacement function. In the case of functions which only produce semantically similar states, that model can effectively be an empty function body such as that which we used for \ic{printf} (shown in Listing \ref{code:short_printf}). These changes may of course impact soundness and completeness, and so it is important to consider them on an application by application basis. 

\begin{lstlisting}[label={code:short_printf},caption={Low fidelity mode for \ic{printf}.},float ]
int printf(const char * format, ...) { return 0; }
\end{lstlisting}

Avoidable code that is responsible for significant state forking also appears within the applications themselves. For example, \ic{tcpdump} makes use of \ic{libpcap} to load packets for processing. It can achieve this by making use of one of two different parsers, which effectively produce the same end result, but one of which contains significantly fewer branches on symbolic data than the other. Listing \ref{code:fread_in_pcap_1} shows the loop which iterates over a list of function pointers, and calls each function with the header from the input file. The function indicates that it can load the data by returning a non-null value. If a null value is returned the next function is consulted. Consequently, we can force the less complex parser to be utilised by inserting a single assumption, as shown in Listing \ref{code:check_headers_in_pcap_2} at line 1.

\begin{lstlisting}[breaklines=true,label={code:fread_in_pcap_1},caption={\ic{fread} in \ic{pcap}.},float]
amt_read = fread((char *)&magic, 1, sizeof(magic), fp);
for (i = 0; i < N_FILE_TYPES; i++) {
    p = (*check_headers[i])(...);
    if (p != NULL) 
        goto found;                                           
}
\end{lstlisting}

\begin{lstlisting}[breaklines=true,label={code:check_headers_in_pcap_2},caption={\ic{check\_headers} in \ic{pcap}.},float]
klee_assume(magic == TCPDUMP_MAGIC);
if (magic != TCPDUMP_MAGIC && ...) {
    return (NULL);  /* nope */      
}  
\end{lstlisting}

\subsubsection{State Forking in Input-Discarding Code}\label{section:state_forking_input_discarding}

Another common source of semantically similar states is code which skips over input bytes while searching for some delimiter or terminating character. Listing~\ref{code:while_loop_over_read_pointless} gives an example from \ic{zlib}, which is intended to read and discard a comment in the input. Owing to the conditional statements, the number of states quadruples on each loop iteration. In~\ic{zlib}, this construct is responsible for 42\% of forked states, while similar constructs appear in several of the other targets; e.g. \ic{jasper} (74\%), \ic{file} (63\%) and \ic{libjpeg} (91\%). 

\begin{lstlisting}[label={code:while_loop_over_read_pointless},caption={A while loop with discarded data in \ic{zlib}.},float]
if ((flags & COMMENT) != 0) {   
    while ((c = get_byte(s)) != 0 && c != EOF);
}
\end{lstlisting}

In each of these targets we resolved the issue via the insertion of assumptions.  The mitigation for the example shown in Listing \ref{code:while_loop_over_read_pointless} can be seen in Listing \ref{code:zlib_assume}.  Assumptions can be seen as guidance through the code by compromising theoretical completeness (which is effectively impossible in practice) for analysis in the more interesting areas of the code, as deemed by a domain expert.

\begin{lstlisting}[label={code:zlib_assume},caption={Avoiding forking via an assumption in \ic{zlib}.},float]
if ((flags & COMMENT) != 0) {   
    c = get_byte(s);
    klee_assume(c == 0 | c == EOF);
    while (c != 0 && c != EOF) ;
}
\end{lstlisting}

\subsection{Resource Consumption by States on Error Paths}
\label{sec:challenge_uninteresting}


Often, once a path has reached a certain point, it will never return to interesting code that we wish to analyse. Usually, this arises when a state has reached an error path and the remainder of its processing exclusively involves standard error handling. In order to avoid analysing such uninteresting paths, we modify the code of the highest function in the error handling chain to immediately exit the program, which terminates the state.


Alternatively, in some situations, the error handling function may only be invoked after expensive processing and it may be preferable to add assumptions which ensure that no path can enter the code that reaches an error state in the first place. The downside of this approach is that each line of code that may reach an error state must be individually handled. Listing~\ref{code:example_of_error_call} gives an example of a blocked path \emph{after} the introduction of a \ic{klee\_assume} statement.

\begin{lstlisting}[label={code:example_of_error_call},caption={A call to \ic{TIFFErrorExt} in \ic{LibTIFF}.}, float]
klee_assume(tif->tif_dir.td_nstrips);
if (!tif->tif_dir.td_nstrips) {
    TIFFErrorExt(...);
    goto bad;
}
\end{lstlisting}
  
Our mitigation poses risks as bugs may be present in the uninteresting paths we have manually sliced out. However, the alternative is fatal state space explosion. We reduced this risk by verifying that the sliced out paths are simple and unlikely to expose bugs.


\subsection{Expensive Initialisation of Static Data}
\label{sec:challenge_expensive_initialisation}


Code related to static instantiation of concrete data can often be expensive due to the sheer number of instructions involved. For example, \ic{libjpeg} contains a function to instantiate an array with static data. This function contains a number of nested loops, that when unrolled amount to 19,791,892 LLVM instructions. Essentially, the flattened loop encompasses 1.2x more instructions than that executed by KLEE in a one hour run for \ic{libjpeg}. Furthermore, the function lies on a path which all states must pass through in order to continue deeper into the program. During our experimental runs, \emph{no} paths managed to break through the function.

% ...

A less extreme case was also encountered during the analysis of \ic{tcpdump} when it initialises a table for the calculation of \texttt{CRC} values.    

There are two ways to resolve issues such as these. In cases like \ic{libjpeg}, where the total number of instructions that must be executed is excessively large, one can statically compute the resulting table ahead-of-time and simply embed it in the target software. This was the approach taken for \ic{libjpeg}, and code coverage increased from 13.18\% to 19.18\%. In cases like \ic{tcpdump}, where the number of instructions is sufficiently small that a single state exercising them is not overly time consuming, then the initialisation code can often be lifted to a point in the program before any state forking has taken place.    

\subsection{Overhead of Initialisation Code}
\label{sec:challenge_overhead_initialisation}


Some applications have distinct subcomponents which can be more efficiently tested in isolation.  For instance, \ic{w3m} contains a library called \ic{libwc}, which is used for character conversion. The library is accessed through a single API, \ic{wc\_Str\_conv\_with\_detect}, and it is invoked to convert the stream of input characters prior to HTML parsing. The input provided to \ic{w3m} is directly passed to this API, with no constraints imposed. Therefore, testing the safety of the character conversion code entails the unnecessary overhead of first analysing the extensive initialisation routines which \ic{w3m} goes through prior to reaching the interesting API. 

To avoid this, a driver program can be constructed which calls the API function of interest. In the case of \ic{w3m}, this is also a convenient way to solve another problem. Depending on the locale information for the machine on which \ic{w3m} is running, and the values of the initial bytes of input, \ic{wc\_Str\_conv\_with\_detect} will invoke the appropriate conversion function. Crucially, since the locale is fixed for a given analysis machine and not treated as symbolic by \KLEE, only a limited subset of the conversion functions is available to be invoked, and therefore tested. In the driver program, shown in Listing \ref{code:libwc_driver} we can simply mark the locale (given by the \ic{hint} variable) as symbolic and \KLEE will discover and test all the available conversion functions.

\begin{minipage}[c]{0.45\textwidth}
\begin{lstlisting}[label={code:libwc_driver},caption={\ic{libwc} driver program.}]
from = WC_CES_UTF_8;
klee_make_symbolic(&hint, sizeof(wc_ces), "hint");
klee_assume(hint == WC_CES_US_ASCII || hint == ...);
to = WC_CES_UTF_8;
s = wc_Str_conv_with_detect(s, &from, hint, to);
\end{lstlisting}
\end{minipage}

\subsection{Inability to Reason about Meta-Properties of Input}
\label{sec:challenge_meta_properties}


Symbolic execution engines generally track the propagation of symbolic data via direct data-flow; e.g.\ if \ic{a} is symbolic and \ic{b = a + 10}, then the engine will represent \ic{b} as a symbolic value derived from \ic{a} and \ic{10} via the addition operator. However, applications often make decisions based on meta-properties of input data (e.g.\ string length), and these meta-properties usually do not have a direct data-flow dependency or transition on the input data. Therefore, the engine is unable to track and reason about the relationship between the input data and meta-properties. 

\paragraph{\ic{tcpdump} out-of-bounds memory access.}
When processing an input file, \ic{libpcap} reads an integer from the file, bounds its value to the range (0, 262144) and then allocates a buffer based on the value. When \KLEE~ detects a symbolic value being passed to \ic{malloc}, it \emph{concretises} that value through a solver query. 

Later, in \ic{tcpdump}, a number of security-relevant bugs arise when certain sequences of code are triggered which increment a pointer beyond the end of a buffer. Whether they are detectable or not depends on what allocation size \KLEE~ previously produced during concretisation. The impact of concretisation is significant; a small value (e.g. 6) results in 4 bugs being found, a very large value (e.g. 262144) results in 3 bugs, while a value of 128 results in 32 bugs.

A general solution to this problem would require adding functionality to \KLEE to support dynamically allocated memory regions with variable bounds. In our case we handled it by detecting when concretisation occurred, inspecting the code which makes use of the produced buffer and adding assumptions on the concretisation size to ensure it is within a sensible range e.g. for \ic{tcpdump} we added a constraint to bound the allocated buffer's size to the range (128, 256).

\paragraph{\ic{zlib} out-of-bounds memory access.}
In \ic{zlib}, the root cause of why \KLEE~ fails to detect CVE-2005-2096 is similarly due to its inability to track and reason about a metaproperty of input. A key variable in triggering the bug is \ic{min}, the value of which is calculated as shown in Listing \ref{code:inflate_table}. Even though an attacker can control the value of \ic{min} by controlling the values found within the \ic{count} array, there is no direct data-flow dependency between them and thus \KLEE~ cannot reason about any conditions which are based on the value of \ic{min}. Essentially, \ic{min} is not considered symbolic.

\begin{minipage}[c]{0.45\textwidth}
\begin{lstlisting}[label={code:inflate_table},caption={Metaproperty calculation in \ic{inflate\_table}.}]
for (min = 1; min <= MAXBITS; min++)
    if (count[min] != 0) break; 
\end{lstlisting}
\end{minipage}

Such a case is synonymous to a bug that is triggered according to the length of a symbolic string, rather than the string’s actual content. There is no direct way to deal with this issue without extensive modifications to the analysis engine. Instead, we hope that by alleviating the other problems mentioned, we enable a sufficient increase in the number of processed paths where one with the desired property is produced and scheduled. 

\subsection{Inefficient Environment Models}
\label{sec:challenge_environment}


Often, real-world programs interact with code provided by the environment (e.g.\ the kernel and third-party libraries). For such code, the analysis engine has the choice of evaluating it as if it were code within the target application, or skipping over the call and mimicking its effects via a model. The former approach is more straightforward, as an accurate model can be difficult to construct. However, we encountered several situations in which analysing the environment-provided code resulted in critical state-space explosion. 

\begin{lstlisting}[label={code:memcpy},caption={Code snippet of \ic{memcpy}.},float]
void *memcpy(void * s1, const void * s2, size_t n) {
    char *r1 = s1;
    const char *r2 = s2;
    while (n) {
        *r1++ = *r2++;
        --n;
    }
    return s1;
}
\end{lstlisting}

For instance, the \ic{memcpy} functionality provided by \ic{uclibc} is shown in Listing \ref{code:memcpy}. State space explosion growth occurs when $n$ is symbolic as the while loop will potentially create $2n$ states on each \ic{memcpy} call. \KLEE also performs queries at three different locations on each iteration of the loop: (1) to check the value of $n$ at the head of the loop, (2) to check that the dereference of $r1$ is in bounds, and (3) to check that the dereference of $r2$ is in bounds. During an analysis run of tcpdump, the while loop was responsible for 50\% of the states produced, and 25\% of the solver queries issued.

The \ic{memset} and \ic{strtol} family of functions have similar issues. A direct mitigation to this problem is to add handlers to \KLEE for such functions, which model their effects without invoking the native code. We were able to alleviate the issues in some situations by adding assumptions which limited the bounds of symbolic variables or concretised them outright.

\section{Evaluation of the Impact of Manual Intervention}
\label{sec:final_results}
We took an iterative approach when introducing our modifications, as detailed in Section \ref{sec:challenges}, and stopped the mitigation process when there was no longer a clear and significant challenge that we can address using manual intervention. As manually mitigating the issues is quite time consuming, for each program we focused on two versions: the head, as provided by version control at the start of our evaluation, and one other version found in \ic{Hemiptera}.

\begin{table*}[t]
\centering
\caption{Coverage and Bug Count when running \KLEE on modified programs}
\label{tab:final_results}
\begin{tabular}{|l|l|r|r|r|r|r|r|r|r|r|}
\hline
\multicolumn{1}{|c|}{\multirow{3}{*}{\textbf{Program}}} & \multicolumn{1}{c|}{\multirow{3}{*}{\textbf{Commit}}} & \multicolumn{3}{c|}{\multirow{2}{*}{\textbf{\begin{tabular}[c]{@{}c@{}}Default State\\ of the Art\end{tabular}}}} & \multicolumn{6}{c|}{\textbf{With modifications}}                                                                                                                                                                                                        \\ \cline{6-11} 
\multicolumn{1}{|c|}{}                                 & \multicolumn{1}{c|}{}                                 & \multicolumn{3}{c|}{}                                                                                             & \multicolumn{2}{c|}{\textbf{Bugs}}    & \multicolumn{2}{c|}{\textbf{H. Bugs}}                                                                                 & \multicolumn{2}{c|}{\textbf{ICov}}                                                                                              \\ \cline{3-11} 
\multicolumn{1}{|c|}{}                                 & \multicolumn{1}{c|}{}                                 & \multicolumn{1}{c|}{\textbf{Bugs}}        & \multicolumn{1}{c|}{\textbf{H. Bugs}}             & \multicolumn{1}{c|}{\textbf{ICov (\%)}}                   & \multicolumn{1}{c|}{\textbf{Mean}} & \multicolumn{1}{c|}{\textbf{$\Delta$}} & \multicolumn{1}{c|}{\textbf{Mean}} & \multicolumn{1}{c|}{\textbf{$\Delta$}} &  \multicolumn{1}{c|}{\textbf{ICov (\%)}} & \multicolumn{1}{c|}{\textbf{$\Delta$ (pp)}} \\ \hline
libjpeg-turbo                                                & a0b7de9*                                              & 0  & -                                                   & 23.1                                                      & 0                                  & 0                                                                           & - & -  & 27.08                                   & +3.98                                      \\ \hline
libjpeg-turbo                                                & 3091354                                               & 1   & 0                                                  & 15.01                                                     & 1                                  & 0                                                                           & 0 & 0 & 21.46                                   & +6.45                                      \\ \thickhline

imginfo                                                & 4212e7e*                                              & 0 & -                                                    & 10.36                                                     & 0                                  & 0                                                                             & - & - & 14.22                                   & +3.86                                                                         \\ \hline

imginfo                                                & b702259                                               & 1 & 1                                                     & 10.39                                                     & 2                                  & +1                                                                         & 2 & +1 & 11.3                                    & +0.91                                      \\ \thickhline

file                                                   & df74b09*                                              & 0  & -                                                   & 12.61                                                     & 0                                  & 0                                                                            & - & - & 16.62                                   & +4.01                                      \\ \hline
file                                                   & b6e8437                                               & 0  & 0                                                   & 13.83                                                     & 1                                  & +1                                                                         & 1 & +1 & 16.25                                   & +2.42                                      \\ \thickhline
tcpdump                                                & 17a3c288*                                             & 32  & -                                                  & 4.94                                                      & 61                                 & +29                                       & - & - & 8.89                                    & +3.95                                      \\ \hline
tcpdump                                                & a9e4211                                               & 40 & 2                                                    & 4.91                                                      & 69                                 & +29                                    & 3 & +1 & 10.03                                   & +5.12                                      \\ \thickhline

zlib                                                   & cacf7f1d*                                             & 0  & -                                                   & 26.7                                                      & 0                                  & 0                                     & - & - & 30.66                                   & +3.96                                      \\ \hline
zlib                                                   & 7c2a874                                               & 0 & 0                                                    & 26.56                                                     & 1                                  & +1                                    & 1 & +1 & 33.83                                   & +7.27                                      \\ \thickhline

tiff2pdf                                               & d57ccfc9*                                             & 0 & -                                                    & 8.48                                                      & 0                                  & 0                                    & - & - & 7.68                                    & -0.80                                      \\ \hline
tiff2pdf                                               & f64949a                                               & 0  & 0                                                   & 8.42                                                      & 0                                  & 0                                     & 0 & 0 & 9.17                                    & +0.75                                      \\ \thickhline
flac                                                   & d2cb0d1*                                              & 0 & -                                                    & 12.48                                                     & 0                                  & 0                                    & - & - & 17.3                                    & +4.82                                      \\ \hline
flac                                                   & d8d1717                                               & 0 & 0                                                    & 12.29                                                     & 1                                  & +1                                    & 1 & +1 & 17.3                                    & +5.01                                      \\ \thickhline

w3m                                                    & 1ac245b*                                              & 0  & -                                                   & 9.01                                                      & 0                                  & 0                                    & - & - & 9.43                                    & +0.42                                      \\ \hline
w3m                                                    & 02ba3d6                                               & 4 & 0                                                    & 9.07                                                      & 5                                  & +1                                    & 0 & 0 & 9.13                                    & +0.06                                      \\ \thickhline
w3m-driver                                             & 1ac245b*                                               & 0  & -                                                   & 9.01                                                      & 0                                  & 0                                    & - & - & 10.13                                   & +1.12                                      \\ \hline
w3m-driver                                             & 02ba3d6                                               & 4  & 0                                                   & 9.07                                                      & 24                                 & +20                                   & 1 & +1  & 11.13                                   & +2.06                                      \\ \hline \hline
\multicolumn{10}{|r|}{Average change}                                                                                                                                                                                                                                                                                                                                                                 & +3.08                                      \\ \hline
\end{tabular}
\end{table*}

Table \ref{tab:final_results} shows the results obtained by running \KLEE on the target applications, using manual intervention to alleviate issues, as described in Section \ref{sec:challenges}. 

\subsection{Results}

\paragraph{Code Coverage}
Our modifications resulted in increased coverage in every analysed program, except one version of \ic{tiff2pdf}. The most significant improvements in percentage points (pp) were in \ic{zlib-7c2a874} (+7.27), \ic{libjpeg-3091354} (+6.45) and \ic{tcpdump-a9e4211} (+5.12). The average increase in coverage across all applications and versions was 3.08 percentage points.

Regarding \ic{tiff2pdf}: despite our attempts, \KLEE still struggled to get beyond the initial portion of the parser. The parser contains a large number of loops with branches based on symbolic data and there was no clear set of logical assumptions which would help \KLEE progress. The reason for the drop in code coverage is that the added assumptions, which aim to eliminate paths that appear superfluous, e.g. error paths, did not result in a corresponding increase in coverage elsewhere.

\paragraph{Bugs Found}
Without manual intervention, \KLEE found bugs in six of the program versions analysed. With manual intervention, bugs were found in three new programs, namely \ic{file}, \ic{zlib} and \ic{flac}. Within the six program versions where bugs were discovered with manual intervention, the modifications resulted in a increase in the number of bugs found. 

The most significant improvement was in \ic{tcpdump} with 29 new bugs discovered. At the time of analysis, these bugs were still present in the release version. Notable interventions responsible for the increase in bug count in tcpdump include: (1) the adding of assumptions to avoid \KLEE concretising integers to zero that later were used to control allocation sizes, and (2) the replacement of libc's \texttt{printf} with a low fidelity function

The number of bugs found, and their diversity, also increased significantly on \ic{w3m-02ba3d6}, from 4 bugs across 2 conversion functions to 24 bugs across 10 conversion functions. The most significant change for w3m was the use of a driver program, which enabled the direct analysis of compartmentalized libraries without the need to go through HTML parsing code. 

\paragraph{Required Effort} Scientifically quantifying manual effort is always going to be a tricky endeavour. It depends on numerous variables where their measurements are often subjective, e.g. the skill of the manual analysts and the complexity of the application. Nevertheless, the time required for indentifying and making the changes was a matter of minutes, i.e.\ less that 15 minutes. In total we spent approximately two days per target (16 hours), which includes the analysis time taken by \KLEE.

We conclude that the above data validates the assumption that manual intervention \emph{can} alleviate a variety of issues encountered when analysing real world programs.

\subsection{Limitations}

Manual intervention resulted in an increase in code coverage across most targets, and an increase in bugs found across almost half of the targets. However, the number of bugs found did not increase for 10 of the 18 targets on which we applied manual intervention. Despite increasing code coverage for many of these 10 targets we were still not alleviating a sufficient number of problems encountered by \KLEE.

Manual intervention appears to hit diminishing returns after a few iterations of our process. In general, we observed that at first there are a very small number of problematic constructs that when addressed result in significant gains. Over time state forking and solver query time (which are usually the limiting factors) become more evenly distributed across the covered code and it becomes more difficult to determine what paths can be eliminated, and which ones should be kept. The cost of discovering the correct assumptions increases, and the pay-off, in terms of additional code coverage, decreases.

To give a concrete example: after 7 iterations of changes and analysis on \ic{file}, 46\% of all forks were distributed across its \ic{parse} function. It is a complex function through which all states must pass, and thus it could not be replaced with a low fidelity model. As no paths were obviously uninteresting we also could not utilise assumptions to remove them. 

Predicting the impact and validity of a change can also be quite difficult. On a number of occasions, we added an assumption only to discover at runtime that it was incompatible with an assumption added elsewhere, or conditions imposed by the source code itself. 

For instance, Listing \ref{code:w3m-code1} is a code-snippet taken from \ic{w3m}, where line 3 is responsible for a large number of forks (45\%). The first attempt made to alleviate this issue involved the addition of the \ic{klee\_assume} shown on line 2. However, the assumption is invalid, as states which reach this code have passed through functions that guarantee that the string is NULL terminated. Such a mistake was only realised during runtime as \KLEE printed a warning regarding the logical inconsistency.

 \begin{lstlisting}[label={code:w3m-code1},caption={An insertion of an invalid assumption for \ic{w3m}.}]
for (i = 0; i < s->length; i++) {
    klee_assume(s->ptr[i] != '\0');
    if (s->ptr[i] == '\0') {...}
 \end{lstlisting}

\section{Threats to Validity}
\label{sec:threats_to_validity}
\paragraph{Analysis Validity.} The key threat to validity concerns the quality of our modifications, which are derived manually. Whilst we were cautious, a priori knowledge of the bugs present in \ic{Hemiptera} could have indirectly influenced us during our experiments. We reduced this threat in two ways.  Firstly, apart from buggy commits, we also experimented with the head versions of the programs where no bugs were known. Secondly, modifications are justified according to concrete evidence; they are strictly included only if they address issues that are  apparent in the data collected from experimental runs.

It is possible that different symbolic execution engines would have presented different challenges that were not apparent within \KLEE. We primarily selected \KLEE as it is seen as the state-of-the-art and has been employed extensively in previous work \cite{cadar2011symbolic,marinescu2012make,ramos2015under,cui2013verifying}.  

\paragraph{Validity of \ic{Hemiptera}.}
A concern associated with internal validity of \ic{Hemiptera} is the potential for false positives in the bugs included. We minimised this threat by triggering all bugs to confirm their presence. Another potential concern is biases in the selection process. Although completeness (i.e. recording \emph{all} previous bugs in an application) was never our objective, we did strive to consider the majority of bugs in a practical manner so that such biases may be reduced. 

Another potential threat is the generalisability of \texttt{Hemiptera} with respect to other applications. We address this concern by considering a reasonably high number of bugs across a diverse set of applications.

\section{Related Work}
\label{sec:related_work}
In this section, we give an account on previous works aimed at making symbolic execution more effective for bug finding. We also give a short review on previously proposed bug datasets, and contrast them to \ic{Hemiptera}.

\subsection{Symbolic Execution}

There is a substantial amount of research that has sought to alleviate the limitations of path-based symbolic execution. Many of these approaches involve applying symbolic execution on fragments of the target applications. For instance, \ic{ZESTI}~\cite{marinescu2012make} uses \KLEE to analyse program paths, exercised by regression tests, against all possible inputs, while \ic{UC-KLEE}~\cite{ramos2015under} is an extension of \KLEE capable of analysing arbitrary function code with under-constrained symbolic input. 

State pruning~\cite{bugrara2013redundant} has been employed to remove redundant states. For instance, AEG~\cite{avgerinos2011aeg} only considers states that satisfy a given precondition, whilst Yi et al.~\cite{yi2017eliminating} propose pruning off states that have their path conditions subsumed by previous exploration. FIE~\cite{davidson2013fie} uses state pruning to scale symbolic execution for bug discovery in firmware programs. It~also makes loop counters symbolic in order to limit the number of states forked at looping constructs.

Concolic execution has been successful in avoiding some of the issues of symbolic execution when applied to whole applications~\cite{cha2012unleashing, stephens2016driller, godefroid2012sage}. It is of interest to assess manual intervention done within the context of concolic exection, but such an endeavour would warrant a full study on its own. Mayhem~\cite{cha2012unleashing} takes a hybrid approach; it begins with symbolic execution, with forking enabled, but switches to concolic execution once a threshold is reached in order to avoid excessive memory consumption. Moreover, in both concolic and symbolic execution, there has also been research aimed at automatically resolving state space explosion due to loops and frequently called functions. Christakis et al.~\cite{anipaper} make use of summarisation to verify the memory safety of the ANI Windows Image Parser. A~number of researchers have presented systems that make use of state merging~\cite{avgerinos2014enhancing, kuznetsov2012efficient}. 

As detailed in Section~\ref{sec:challenge_meta_properties}, indirect data flows pose challenges to tracking symbolic variables. Corin et al.~\cite{corin2012taint} employ control-flow analysis techniques to mitigate this. In relation to the challenge in Section \ref{sec:challenge_uninteresting}, Rawat et al.~\cite{rawat2017vuzzer} propose heuristics to automatically identify error paths to avoid their consideration.

We expect that many of the mitigations we have documented, and manual intervention in general, would be beneficial in allowing the above tools to gain increased code coverage and bug finding capabilities, but this remains to be evaluated empirically.

\subsection{Bug Datasets}

One of the earliest works related to bug datasets was carried out by Wilander et al. \cite{wilander2002comparison}. The authors manually developed 44 C function calls to evaluate tools designed for static intrusion prevention. Zitser et al. \cite{zitser2004testing} presented a dataset comprised of 14 bugs based on exploitable buffer overflows found in three open-source applications. The bugs were extracted and transformed into smaller program models to enable tools, which suffer from scalability issues, to be evaluated. Kretkiewicz et al. \cite{kratkiewicz2005using} built upon the work of Zitser et al. by proposing a greater number of test cases (as opposed to just 14). A similar approach was taken by Ku et al. \cite{ku2007buffer}, where their dataset consists of 298 generated test cases (half of which are faulty) derived from 22 bug kernels. Overall, these datasets have one underlying issue; their test cases are all synthetic. Threats to external generalisability arise as they fail to capture significantly the complexities of real-world software. 

BugBench \cite{lu2005bugbench} was one of the first notable bug dataset to consider entire real-world applications, rather than bug kernels. Unfortunately, the main limiting factor of BugBench is its relatively low bug count (19 bugs in total), which hinders its diversity. BegBunch \cite{cifuentes2009begbunch} was later released. Its test cases were generated by making use of bug kernels extracted from real-world software, similar to the approach taken by Zitser et al. \cite{zitser2004testing}. However, a greater number of kernels were used, namely 64. In total, their relevant corpus includes 181 synthetic test cases. 

The Juliet Test Suite \cite{boland2012juliet} is a large collection of 81,056 synthetic C/C++ and Java programs provided by NIST Software Assurance Metrics and Tool Evaluation (SAMATE). All together, the test cases encompass 118 different CWEs. Furthermore, Shiraishi et al. \cite{shiraishi2015test} present a corpus of 1,276 simple synthetic programs that encapsulate common characteristics of automotive software. Only half of the test cases contain flaws.

More recently, Dolan-Gavitt et al. \cite{dolan2016lava} proposed LAVA, a technique that automatically injects vulnerabilities into source code. The authors made use of their tool to inject vulnerabilities into four real-world applications; for example, LAVA inserted 1064 bugs in the \texttt{readelf} application. Although LAVA facilitates the construction of a large bug corpus, there is no guarantee that the injected flaws are representative of issues that occur in real world code. 

The DARPA Cyber Grand Challenge \cite{song2015darpa,song2016darpa} provided a set of challenges comprised of 131 vulnerable binaries, constructed for the competition. They are designed to run on DECREE, a custom operating system which supports a limited number of system calls and was also specifically created for the challenge. As with the other synthetic corpora, it is not clear how well the programs, their vulnerabilities and execution environment correspond with real software. 

\ic{Hemiptera} has more real bugs than any other benchmark to date. Although it still faces threats related to generalisability, a researcher should have a higher degree of confidence when using Hemiptera, as results correlate with a large number of real-world bugs.

\section{Conclusion}
\label{sec:conclusion}
In this work, we present \ic{Hemiptera}, a novel bug suite, and use it to investigate the issues faced by \KLEE, a state of the art symbolic execution engine, when analysing real-world software. We document an in-depth analysis of the issues that hinder \KLEE and categorise the main challenges faced. While \KLEE is capable of finding bugs on some of the targets in \ic{Hemiptera}, our evaluation highlights the fact that a number of commonly occurring code constructs can severely hinder its progress. 

We utilised the above categorisation to experimentally evaluate the effectiveness and limitations of manual intervention in symbolic execution. Whilst the approach is not without its issues, our experience shows an improvement in both bugs found and code coverage. According to these results, we conclude that manual intervention is likely to be a rewarding effort in the real world deployment of symbolic execution tools. We believe that the analysis of the issues found should also prove useful for future work on automated solutions, and that \ic{Hemiptera} will be a valuable resource in evaluating such solutions.

\bibliographystyle{plain}
\bibliography{mybib}

\end{document}